\documentclass[]{interact}
\usepackage{epstopdf}
\usepackage[caption=false]{subfig}

\usepackage[numbers,sort&compress,merge]{natbib}
\bibpunct[, ]{[}{]}{,}{n}{,}{,}
\renewcommand\bibfont{\fontsize{10}{12}\selectfont}

\newcommand\apj{Astrophys. Journal }
\newcommand\apjs{Astrophys. Journal Supplement }
\newcommand\aap{Astron. Astrophys. }
\newcommand\pra{Phys. Rev. A }
\newcommand\prl{Phys. Rev. Letters }
\newcommand\jcp{Journal of Chemical Physics }
\newcommand\jqsrt{J. Quant. Spectr. \& Rad.Transfer }

\theoremstyle{plain}

\theoremstyle{definition}

\theoremstyle{remark}

\begin{document}


\title{The quasi bound spectrum of H$_2$ }

\author{
\name{E.~M. Roueff\textsuperscript{a}\thanks{CONTACT E.~M. Roueff. Email: evelyne.roueff@obspm.fr \\
Molecular Physics  https://www.tandfonline.com/doi/10.1080/00268976.2022.2122887 } and H. Abgrall\textsuperscript{a}}
\affil{\textsuperscript{a}Observatoire de Paris, 5, Place J. Janssen, Meudon 92190, France}
}

\maketitle

\begin{abstract}
We compute the radiative ro-vibrational emission spectrum of H$_2$ involving quasibound states via a simple numerical method of resolution of the Schr{\"o}dinger equation by introducing a modified effective molecular potential. The comparison of the eigenvalues obtained with our approximation and other theoretical methods based on scattering resonance properties is excellent. 
Electric quadrupole and magnetic dipole contributions are calculated and we confirm the previous computations of  Forrey of the electric quadrupole transition Einstein coefficients. The astrophysical relevance of such quasibound levels is emphasized.
\end{abstract}


\section{Introduction}

The presence of  quasibound levels  of molecular H$_2$ located above its dissociation limit is known since many years from its VUV emission spectrum,
 as reported by Dabrowski \cite{Dabrowski:84} and Roncin \& Launay \cite{Roncin:94}. 

The first tentative computations of quasibound levels have been performed  by Waech and Bernstein \cite{waech:67} by studying the energy dependence of the phase-shifts in an analytic expansion of the H$_2$ ground state potential.
Schwenke \cite{Schwenke:88}, followed by Selg \cite{Selg:12}, have subsequently developed theoretical methods to compute the position and dissociation widths of the different quasibound levels of H$_2$ with the available numerical potential energy functions of H$_2$ from the properties of the scattering matrix.

These levels, also defined as shape resonances \cite{Bain:72}, have however attracted little interest until 2016 in the astrophysical community, where Forrey \cite{Forrey:16} computed the enhancement
in the  H + H $\rightarrow$ H$_2$ + $h \nu$ electric quadrupole radiative association reaction provided by these resonances  and its potential relevance for the primordial universe chemistry. Forrey \cite{Forrey:16} also computed the Einstein emission coefficients of the electric quadrupole transitions involving quasibound - bound and quasibound - quasibound levels.
In the same year, Pike et al. \cite{Pike:16} report several highly excited H$_2$  transitions observed in Herbig-Haro 7, including the (2-1)S(27) transition where
the upper level lies above the dissociation limit, corresponding thus to an upper quasibound level,  but no specific comment to this issue is discussed and the value of the corresponding Einstein coefficient is missing
without any further remark.

The computations presented by Forrey \cite{Forrey:16} were subsequently vigorously criticized by P\'erez et al. \cite{Perez:19} who claim for orders of magnitude different values 
of the transition and dissociative probabililities. These last results have been reasserted by Molano and Arango \cite{Molano:21} who discuss phase-space propagation and stability analysis of the 1-dimensional Schr{\"o}dinger equation for finding bound and resonance states of rotationally excited H$_{2}$.

It is worth mentioning at this point that the Ubachs group in Amsterdam has been able to probe some of these highly excited rotational quasibound levels thanks to
a new experimental setup involving three VUV lasers \citep{Lai:21,Lai:21b} , where  highly rotationally excited H$_2$ is produced from the photodissociation of H$_2$S induced by a 2-photon absorption experiment. These studies beautifully confirm  the energy level positions of H$_2$ derived from the 
theoretical computations of  the ground state electronic potential of H$_2$ by Czachorowski et al. \cite{Komasa:18}. 

We revisit in this paper the infrared spectrum of H$_2$ involving its quasibound levels and provide the associated transition wavenumbers and corresponding Einstein emission coefficients. We introduce both the contribution of electric quadrupole and magnetic dipole moments as first emphasized by Pachucki and Komasa
\cite{Pachucki:11} and introduced by Roueff et al. \cite{Roueff:19} for the bound-bound infrared spectrum of H$_2$.
We present in Section 2 the numerical method and the molecular data used in the computations and  report our results in Section 3. Our summary and conclusions are given in Section 4.

\section{Numerical computations}
We are mainly interested in the electric quadrupole and magnetic dipole transition matrix elements that allow to quantify the possible transition intensities
and take advantage of the previously computed quasibound energy level positions \citep{Schwenke:88, Selg:12}. 

\begin{figure}[h]
\centering
{%
\resizebox*{11cm}{!}{\includegraphics{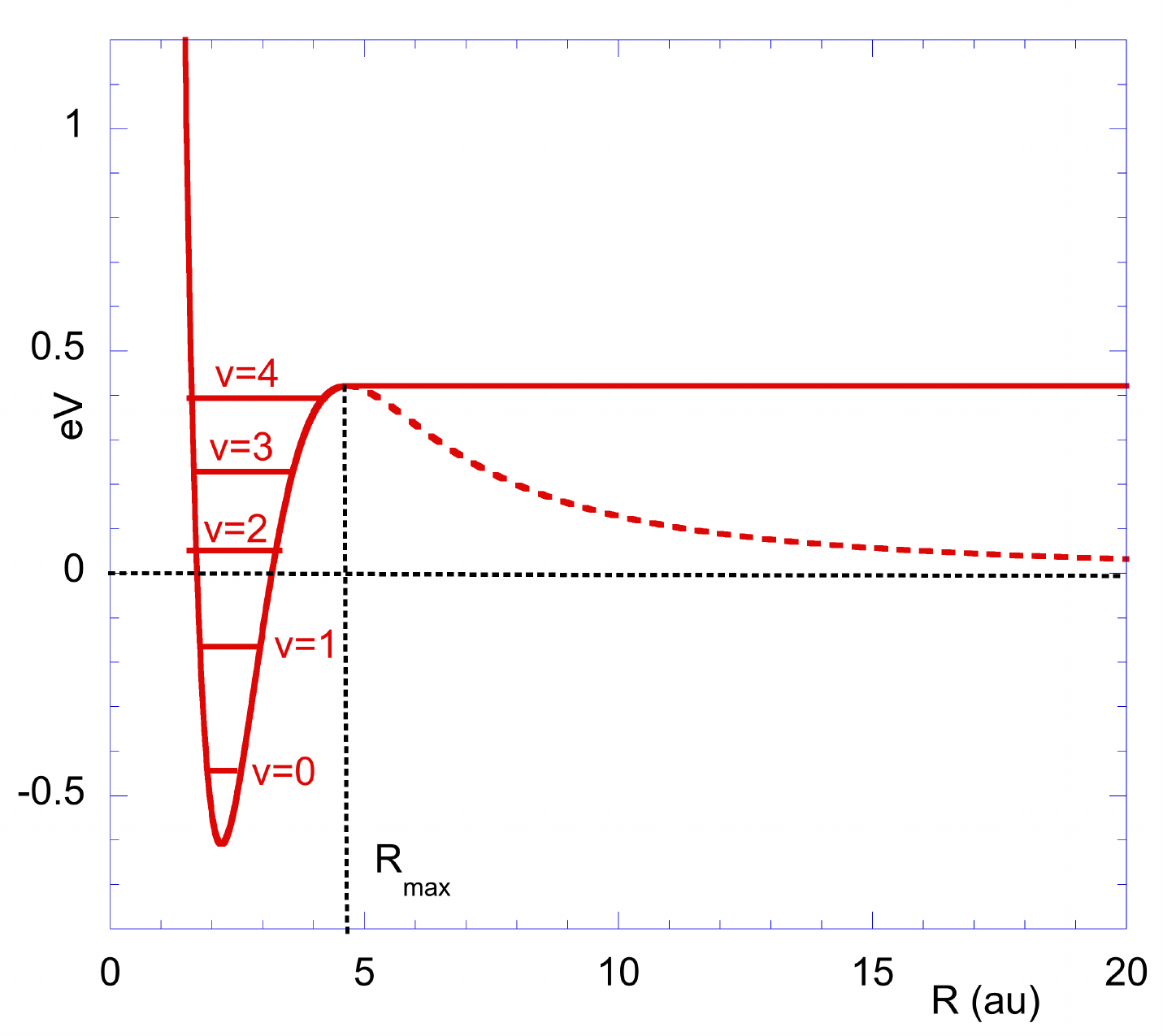}}}\hspace{5pt}
\caption{Modified effective potential of H$_2$ for J=29, $V_{eff}^{mod}$, displayed as a full line and corresponding eigenvalues. $V_{eff}(R)$ is displayed as a dotted line from $R=R_{max}$.} \label{fig:potential}
\end{figure}

The
procedure is the following:
we extend, from the value $R = R_{max}$, the effective potential energy function  
\begin{equation}
V_{eff}(R,J) = V(R) + \frac{\hbar^2 J\,(J+1)}{2 \mu R^2} 
\end{equation} 
by a  constant value equal to the maximum of the bump induced by the centrifugal barrier as displayed by the solid curve in Fig \ref{fig:potential}, that is defined as  $V_{eff}^{mod}(R,J)$ :
\begin{eqnarray}
 V_{eff}^{mod}(R,J) & = &V(R) + \frac{\hbar^2 J\,(J+1)}{2 \mu R^2} \,\,\,\,\,\,    {\mbox{${if   R    \leq  R_{max}} $}} \\
                 &=  &  V (R_{max})  \,\,\,\,\,\,\,\,\,\,\,\,\,\,\,\,\,\, \,\,\,\,\,\,\,\,\,\,\,\,\,\,\,\,\,\,   {\mbox{${if   R   \geq R_{max}}$ }},   \nonumber
\end{eqnarray}
The potential energy function $V(R)$ is the adiabatic potential of the X ground state of H$_2$ where the 
adiabatic correction computed  in \cite{Pachucki:14} is appended to the  Born-Oppenheimer interaction energy displayed in \cite{Pachucki:10}. We then solve the  Schr{\"o}dinger equation to find the bound levels within this potential using our standard renormalized Numerov method \citep{johnson:77}:

 \begin{equation}
 \label{eq:schr}
- \frac{\hbar^2}{2 \mu}  \cdot  \frac{d^2f_{v,J}(R)}{dR^2}  + V_{eff}^{mod}(R,J) f_{v,J}(R) =  E_{v,J} f_{v,J}(R) 
\end{equation} 
where $\mu=M_\mathrm{p}/2$ is the nuclear reduced mass of H$_2$,

The step size in the radial variable is 0.01 au (atomic unit). We have verified that reducing that value by a factor of two does not change the first four numerical 
figures of the radiative emission rate. The maximum range over which the wave-functions are propagated is 12 au. The reduced mass is taken as 918.0764 $m_e$\footnote{A more acccurate value is now available from CODATA21 \cite{CODATA21} but we keep the same value as in our previous calculations \cite{Abgrall:94} to preserve the consistency of our results.}.
The solutions found above the dissociation limit at $V(R)=0$ correspond to the quasibound solutions. 

In the special case where J=29, corresponding to Figure \ref{fig:potential}, we obtain two bound solutions corresponding to $v$ = 0 and 1 and three quasibound levels
corresponding to $v$ = 2, 3 and 4. The shape of the potential beyond the bump corresponding to $R_{max}$ has in fact no influence on the search of bound eigenvalues, as checked here for the positions of levels corresponding to $v  =0 $ and 1.
Table \ref{tab:J29} displays the eigenvalues of the  Schr{\"o}dinger  equation corresponding to J=29  obtained with  $V_{eff}$   and  $V_{eff}^{mod}$. We also report the values computed by Schwenke \cite{Schwenke:88} with his adiabatic potential for comparison in order to test our method within the closest approximation (see below). The bound energies are identical with the two procedures. The comparison between columns 3 and 4  concerning the quasibound levels (with positive eigenvalues) 
is  very impressive and the differences are less than a wavenumber, except for the highest $v$, that is close to the maximum located at 3403.209507 cm$^{-1}$ (corresponding to 
0.4219442eV)   for J=29. 

\begin{table} [h]
\tbl{Eigenvalues of molecular H$_2$ in its ground electronic state corresponding to J=29.}
{\begin{tabular}{lccc}  \toprule
$v$ &  $V_{eff}$ & $V_{eff}^{mod}$  &  reported in  \cite{Schwenke:88}  \\  
   &   cm$^{-1}$   &    cm$^{-1}$ &  cm$^{-1}$  \\
  \hline
0       &     -3693.15    &     -3693.15   &      \\  
1      &      -1471.81   &       -1471.81  &  \\  
2       &    ---     &     473.39   &    474.14    \\  
3       &     ---      &   2092.35   &  2092.99 \\  
4       &     ---      &     3252.73  &  3251.01 \\  
  \hline
  \end{tabular}}
\label{tab:J29}
\end{table}

We compare then in Table \ref{tab:energy} the quasibound eigenvalues obtained with our procedure (E$_{adiab}^{pw}$) and the results reported in \cite{Schwenke:88,Selg:12}, where the authors determine the resonance parameters (position and width) from the poles of the scattering matrix at complex energies \cite{Schwenke:88} or by the detailed investigation of the energy dependence of the phase shifts and amplitude of the real stationary scattering-state function \cite{Selg:12}. Schwenke \cite{Schwenke:88} uses the Born-Oppenheimer potential function of Kolos and Wolniewicz \cite{Kolos:65}  (E$_{adiab}$)  and additional nonadiabatic correction (E$_{nonadiab}$) whereas Selg \cite{Selg:12} includes the more recent calculations of Komasa et al. \cite{Komasa:11} (E$_{Selg}$). 
Table \ref{tab:energy} also displays the available experimental determinations E$_{exp}$ derived  from the VUV spectra,
as reported from \cite{Selg:12}. 
The values followed by $\dag$ are the results obtained in Lai et al. \cite{Lai:22} from the previously quoted three VUV laser experiment performed in Amsterdam. 

The origin is taken at the dissociation limit. The eigenvalues obtained by using our approximate method, using the most recent  adiabatic potential function, are within less than a wavenumber close to the adiabatic computations of Schwenke \cite{Schwenke:88}. Such an agreement fully corroborates  our approach, as the  adiabatic potential functions used in the two computations are very close. 

\begin{table} [h]
\tbl{Comparison between different methods to compute the eigenvalues of quasibound levels of H$_2$ and experimental derivations.}
{\begin{tabular}{lccccccc} \toprule
 &  & E$_{adiab}$  & E$_{nonadiab}$  &  E$_{Selg}$ &  E$_{exp}$   &  E$_{adiab}^{pw}$ & E$_{used}$ \\  
  Ref          &         & \cite{Schwenke:88}&  \cite{Schwenke:88}&\cite{Selg:12}&  &present work        &             \\  
  \hline
 $v_u$ & $J_u$   &   cm$^{-1}$ & cm$^{-1}$ & cm$^{-1}$ &cm$^{-1}$ &        cm$^{-1}$          &  cm$^{-1}$       \\
     \hline
  0 &32 &  432.68 &  428.28      &  ... &         & 431.82 &   428.28    \\
  1 &31 & 1061.27& 1056.79   &  ... &              &1060.47   &1056.79  \\
  2 &29 &    474.14 & 469.65  &  469.81&    470.09      & 473.39 &469.81   \\
   5 & 24&  233.77&   229.44   & 229.32   &    229.11     &  233.16  &  229.32  \\
   3 & 29   &  2092.99 &  2088.70 & 2088.48 &    2088.59     &  2092.35     & \\
   4 & 29 & 3251.01  & 3247.77   &3247.69&   3248.60        &3252.73  & \\
   3 & 28 &  1062.40 &  1057.98  & 1057.94&  ...      &1061.71  &   \\
  4 & 28 &2459.34   &  2455.46  & 2455.00&    ...    &2459.09  &   \\
 4 & 27 &  1561.58 &   1557.38 &   1557.13&  1556.93        &1560.98 &   \\
 4 & 26 &  600.44 &  596.06  &  595.98 &    596.06     & 599.80  &   \\
 5 & 26 & 1948.26  &  1944.53  & 1944.08 &    1943.99     &  1948.02  &   \\
 5 & 25 &  1121.73 &  1117.63  & 1117.36 &   1117.37       &1121.17  &   \\
 6 & 24 &  1523.54 &  1519.98 & 1519.53 &   1519.45      & 1523.33   &   \\
 6 & 23& 772.07  &  768.08  &  767.79&   767.81       & 771.54  &   \\
 7 & 22& 1181.63  &   1178.28 &  1177.83&    1177.76      &1181.49  &   \\
 7 & 21& 510.08  & 506.26   & 505.93&  505.9310$\dag$    & 509.60   &   \\
 8& 20&   917.19& 914.11   &   913.67&   913.76       &917.20 &   \\
 8 & 19&   331.40&   327.78 & 327.43 &   327.4291$\dag$   & 330.96  &   \\
 9 & 18&   721.83  &  719.11   & 718.69 &   718.67          &722.24 \\
 9& 17&  228.65 & 225.31   & 224.95&    224.9410$\dag$     & 228.25  &   \\
 10 & 16&  581.81 & 579.61   & 579.25 &  ...    &583.29  &   \\
10 & 15&  189.78 &  186.83  &186.46 &   186.4542$\dag$   &189.45  &   \\
11& 14& 475.78  &  474.24  & 474.79 &   475.33      &477.43 &   \\
11 & 13&  195.25 &  192.84  &192.50 &   192.4945$\dag$  & 195.08 &   \\
\hline
  \end{tabular}}
\tabnote{ 
 E$_{exp}$
are the experimental values deduced from the VUV emission spectra \cite{Dabrowski:84} and reported in \cite{Selg:12}, except for the values with a $\dag$ that
are taken from the 3 VUV laser experiment \cite{Lai:22}. E$_{adiab}^{pw}$ is the value derived with our approximation of the modified potential displayed in Fig \ref{fig:potential}. E$_{used}$ is the value used for deriving the transition wavenumbers  displayed in Table \ref{tab:res}.  
}
\label{tab:energy}
\end{table}

We use then the   $f_{v',J'}(R)$ eigenfunctions, solutions of the radial Schr{\"o}dinger equation (\ref{eq:schr}), 
to compute the matrix elements of the electric quadrupole and magnetic dipole  moments within the X ground electronic potential of H$_2$, as in \cite{Roueff:19}, and derive the corresponding Einstein emission coefficients. 

The electric quadrupole emission rate is given by
\begin{eqnarray}
W_{v'J' \rightarrow v"J"}^{EQ} &=& 2.797 \times 10^{-23} \cdot \tilde{\nu}^5  \cdot \frac{1}{(2J'+1)}  \\
                  &      & \cdot \sum_{MM'} |<J'M' f_{X,v',J'}(R) |Q(R)_{Wol}| J"M" f_{X,v",J"}(R) >|_{au}^2 \nonumber
\end{eqnarray}
and the magnetic dipole emission rate is obtained from:
\begin{equation}
W_{v'J \rightarrow v"J}^{MD}= 8.00 \times 10^{-18} \cdot  \tilde{\nu}^3 \cdot J (J+1)  \cdot |< f_{X,v',J}(R) | g(R)|  f_{X,v",J}(R) >|^2
\end{equation}

The electric quadrupole moment $Q(r)_{Wol}$  and   the magnetic dipole moment $g(r)$ functions are  taken respectively from \cite{wolniewicz:98}  and
 \cite{Pachucki:11} and expressed in atomic units. 
The transition wavenumbers $\tilde{\nu}$ introduced in  formulae (4) and (5) are computed from the quasibound energy terms reported in column 8 of Table \ref{tab:energy} that correspond to the most accurate values and are expressed in reciprocal centimeters. 

\section{Results}

\subsection{Potentially observable radiative transitions involving quasibound levels}
\begin{table} [h]
\tbl{Properties of the infrared spectrum emitted from quasibound levels of H$_2$.}
{\begin{tabular}{lccccccccccc} \toprule
Transition& $\tilde{\nu}$ & $\tilde{\nu}^{adiab}$ & $\lambda$ & $\lambda^{adiab}$ & $W^{EQ}$  & $W^{EQ}_{Forrey}$ &    $W^{MD}$    & A$_t$    & A$_d$ & E$_k$ \\  
 label          &   cm$^{-1}$    &      cm$^{-1}$        &      $\mu$m         & $\mu$m    &     s$^{-1}$ &     s$^{-1}$ &   s$^{-1}$ &    s$^{-1}$ &  s$^{-1}$ &cm$^{-1}$  \\
 \hline
  (0-0) S(30)    &   2721.56  &  2722.44& 3.6744  & 3.6732 &  4.727E-06  & 4.73E-06& 0.000E+00  &  4.727E-06 &8.118E-23& 428.28\\
  (0-1) S(30)  &   620.50   & 620.79 &  16.1160  &  16.1085&   3.601E-10  & 3.60E-10&  0.000E+00   & 3.601E-10&   8.118E-23& 428.28 \\
    (1-0) Q(31) &  1974.02  &   1974.82 &  5.0658  &  5.0637 &  3.604E-09   &   3.60E-09 &  2.416E-07  &  5.219E-06  &2.449E-07 &1056.79 \\
   (1-0) S(29)&    4752.38  &   4753.62&  2.1042   &   2.1037& 2.101E-06&  2.10E-06          &    0.000E+00 &   5.219E-06  &2.449E-07  &1056.79 \\
  (1-1) S(29) &  2531.65    &  2532.29& 3.9500  &   3.9490 &   2.872E-06  &    2.87E-06         &  0.000E+00   &   5.219E-06  &2.449E-07   &1056.79\\   
  (1-2) S(29) &    586.98   &  587.09 & 17.0364  & 17.0332&   4.963E-10  &  4.96E-10   &   0.000E+00 &   5.219E-06  &2.449E-07   &1056.79\\  
    (2-0) O(31) &  1387.04  &  1387.73 &   7.2096  &  7.2060&   2.704E-11 &   2.70E-11       &  0.000E+00 & 5.482E-06  &  1.230E-13 & 469.81\\ 
    (2-0) Q(29) &   4165.40 &  4166.53  &   2.4007 &  2.4001  &  3.144E-08  &    3.14E-8     &  1.448E-08  &  5.482E-06  &  1.230E-13 & 469.81\\ 
  (2-0) S(27) &  7034.45 & 7035.97 &    1.4216  &  1.4213 & 2.240E-07 &    2.24E-7     &  0.000E+00  &  5.482E-06  &  1.230E-13 &469.81\\ 
  (2-1) Q(29) &  1944.67  &  1945.20 &  5.1423&  5.1409 &  5.139E-09  &  5.15E-9      &   3.896E-07  &  5.482E-06  &  1.230E-13 &469.81\\ 
 (2-1) S(27) &   4590.29  &  4591.18 & 2.1785 &  2.1781  &  2.944E-06  &    2.95E-6   &  0.000E+00 & 5.482E-06  &  1.230E-13& 469.81 \\ 
(2-2) S(27)&    2399.19  &    2399.54 &  4.1681 & 4.1675  & 1.873E-06 &   1.87E-6      &   0.000E+00 &   5.482E-06  &  1.230E-13 &469.81\\ 
 (2-3) S(27) &  489.60 &  489.53  &  20.4248 &   20.4279 & 2.582E-10 &    2.58E-10     &   0.000E+00 &   5.482E-06  &  1.230E-13 &469.81\\
  (5 -0) O(26) &    8253.69  & 8255.664 &   1.2116&  1.2113 &  4.366E-10 &   4.36E-10  & 0.000E+00  &4.226E-06   & 1.354E-10 &229.32 \\
 (5 -0) Q(24) &  11205.89  &  11208.21&   0.8924 &   0.8922 & 1.005E-09  & 1.01E-10&  1.742E-11   &4.226E-06 & 1.354E-10 &229.32 \\
 (5 -0) S(22)   &   14176.89  & 14179.51 &   0.7054  &  0.7052  & 4.462E-09 &  4.46E-9 &  0.000E+00   &4.226E-06   & 1.354E-10&229.32\\
 (5 -1) O(26) &   5704.04   &  5705.38  &  1.7531  & 1.75273  & 3.225E-09  & 3.23E-09&  0.000E+00   &4.226E-06   & 1.354E-10  &229.32 \\
 (5 -1) Q(24)   &   8455.09 &   8456.74 &  1.1827  &  1.1825  & 1.758E-08 &1.76E-8&  2.025E-10   &4.226E-06   & 1.354E-10   &229.32\\
 (5 -1) S(22) &    11236.10  &  11238.042  &   0.8900  &  0.8898  & 1.141E-08 &1.14E-8&   0.000E+00   &4.226E-06   & 1.354E-10  &229.32 \\
 (5 -2) O(26)  &    3399.86  &3400.63  & 2.9413  &  2.94063 &   9.871E-10  &9.83E-10  & 0.000E+00   &4.226E-06   & 1.354E-10  &229.32 \\
 (5 -2) Q(24)  &   5939.36  & 5940.42 & 1.6837 &   1.6834  &  1.038E-07 & 1.04E-7    &  3.835E-09   &4.226E-06   & 1.354E-10   &229.32\\
 (5 -2) S(22)  &    8524.34  &  8525.67&     1.1731  &   1.1729 & 1.250E-08 & 1.25E-8   &  0.000E+00   &4.226E-06   & 1.354E-10  &229.32 \\
 (5 -3) O(26)   &   1363.64 &   1363.95 &    7.3333  &    7.3317 &  7.582E-10 & 7.57E-10  &  0.000E+00   &4.226E-06   & 1.354E-10  &229.32 \\
  (5 -3) Q(24)   &   3673.35&  3673.90&    2.7223  &   2.7219 & 1.159E-07 &1.16E-7&   1.000E-07   &4.226E-06   & 1.354E-10   &229.32\\
  (5 -3) S(22)  &    6051.26 &  6052.04 &     1.6525  &   1.6523 & 8.018E-07 &8.02E-7&   0.000E+00   &4.226E-06   & 1.354E-10   &229.32\\
 (5 -4) Q(24)  &    1680.50  &   1680.67  &   5.9506  & 5.9500  &  1.238E-09&  1.24E-9 &   4.252E-07   &4.226E-06   & 1.354E-10   &229.32\\
 (5 -4) S(22) &     3832.39 &  3832.73  &    2.6093 &   2.6091  & 2.269E-06& 2.27E-6    &   0.000E+00   &4.226E-06   & 1.354E-10   &229.32\\
 (5 -5) S(22)  &    1892.38   &   1892.42  &  5.2843 &  5.2842  &  3.527E-07& 3.53E-7 &   0.000E+00   &4.226E-06   & 1.354E-10  &229.32 \\
 (5 -6) S(22)  &     271.92  &  271.91 &   36.776 &  36.777 &  1.658E-11& 1.65E-11 &    0.000E+00   &4.226E-06   & 1.354E-10  &229.32 \\
 \hline
  \end{tabular}}
\tabnote{
$\tilde{\nu}$,  $\tilde{\nu}^{adiab}$  are the transition wavenumbers computed respectively from E$_{used}$, E$^{pw}_{adiab}$ in Table \ref{tab:energy},  whereas $\lambda$, $\lambda^{adiab}$  are the corresponding wavelengths. $W^{EQ}$ is the electric quadrupole contribution to the Einstein radiative emission coefficient from the present computations.  $W^{EQ}_{Forrey}$  is the value reported from \cite{Forrey:16}. $W^{MD}$
is the magnetic dipole contribution to the  Einstein radiative emission  coefficients. A$_t$ and A$_d$ are the total radiative decay rate  and the dissociation rate of the upper level. E$_k$ is the
quasibound upper level energy position above the dissociation limit.}
\label{tab:res}
\end{table}

Whereas the full set of radiative transition emission rates between quasibound - bound and quasibound - quasibound levels is computed and available on request, we only report in Table \ref{tab:res}  the potentially infrared observable transitions involving upper
quasibound levels for which the dissociation rate  A$_d$ is significantly smaller than the total radiative decay rate A$_t$. 
The most accurate transition wavenumbers and associated wavelengths correspond to columns 2 and 4. 
 
 One such transition, (2-1) S(27), has already been reported in the literature \cite{Pike:16}, but the Einstein coefficient was not available, preventing any 
 physical interpretation. 
 We also point out that an unidentified transition reported at 2.1044 $\mu$m in \cite{Pike:16} could correspond to
 our prediction of the (1-0) S(29) transition at 2.1042 $\mu$m that has a relatively strong Einstein coefficient of 2.1 $\times$ 10$^{-6}$ s$^{-1}$ and corresponds to an upper energy level at 1056.79 cm$^{-1}$ above the dissociation limit, equivalent to an energy of 53,489 K above $v=0, J=0$, the ground rovibrational level of H$_2$. 
 
 We do not discuss here the derivation of the dissociation rate A$_d$, nor the associated resonance width $\tilde{\Gamma} = A_d / (2 \pi c) $ and take the values reported in \cite{Selg:12}.
We note that the magnetic dipole contribution may significantly exceed the electric quadrupole, as in the case of the (1-0 ) Q(31), (2-1) Q(29) and (5-4) Q(24) transitions.
The uncertainty linked to the quasibound level positions is emphasized by the comparison between $\tilde{\nu}$,  $\tilde{\nu}^{adiab}$  
and $\lambda$, $\lambda^{adiab}$ and is shown to be very small. We have checked that they have no impact on the transition emission rate values reported in Table \ref{tab:res}.
We also report the values of the electric quadrupole transition emission rates calculated in \cite{Forrey:16} for which the agreement is found excellent, confirming thus their approach and contradicting the claims and the results displayed in \cite{Perez:19,Molano:21}.
We note in passing that the predicted transition (1-2) S(29) at 17.03  $\mu$m takes place between two quasibound levels. The dissociation rate of the lower level $v$ = 2, $J$ = 29 is much less than that of the upper level $v = 1, J=31$. Then the dissociation rate is given by the value corresponding to the upper level.

\subsection{The $v=14, \, J=4$ case}

The $v=14, J=4$ level deserves a special comment as it is not clear presently if this level is above or below the dissociation limit. When using the potential function of \cite{Komasa:18}, solving the Schr{\"o}dinger  equation results in a level that is slightly bound, at a level of ${-0.0274}$ cm$^{-1}$. The combination of various Lyman transition wavenumbers involving $v=14, J=4$ by Selg \cite{Selg:12} does not lead to conclusive results. 
We  display the corresponding infrared emission spectrum arising from $v=14, J=4 $ in  the separate Table \ref{tab:v14}, that could bring an additional experimental possibility to probe the nature of that energy level. The dissociation rate is strictly 0 if the level is bound. 
Selg \cite{Selg:11} suggests that this level becomes quasibound
thanks to the hyperfine structure effect and obtains a width $\Gamma$ of 2.1338 $\times$ 10$^{-18}$ eV, corresponding to a lifetime of 308 s and a dissociation rate of 3.2 $\times$ 10$^{-3}$ s$^{-1}$.
 Whereas this lifetime is large for terrestrial laboratory conditions, it is short at the astrophysical scales. That value is  also claimed to be rather sensitive to the relativistic and quantum electrodynamics corrections \citep{Selg:11} . 
\newpage
\begin{table} [h]
\tbl{Characteristics of the infrared spectrum arising from the level $v=14$ and $J$=4 of H$_2$.}
{\begin{tabular}{lccccccccc} \toprule

Transition&  $\tilde{\nu}^{adiab}$ & $\tilde{\nu}$ & $\lambda^{adiab}$ & $\lambda$ &$W^{EQ}$  & $W^{EQ}_{Forrey}$ &    $W^{EQ}_{\rm{J{\acute{o}}{\acute{z}}wiak}}$ &   $W^{MD}$    \\  
 label          &   cm$^{-1}$    &      cm$^{-1}$        &      $\mu$m        & $\mu$m    &     s$^{-1}$ &     s$^{-1}$ &   s$^{-1}$ &   s$^{-1}$ \\
 \hline
(14-0) O(6)&   33703.15 &    33704.60 &  0.2967   &  0.2967 & 7.675E-13 &  7.12E-13    & 3.949E-13 &  0.000E+00 \\
(14-0) Q(4) &    34949.24 &  34950.76 &  0.2861 &  0.2861 &  1.402E-14  &  1.12E-14&1.141E-14& 1.907E-17 \\
(14-0) S(2)    & 35763.67&  35765.22 &  0.2796  & 0.2796  & 4.301E-13 & 4.25E-13  & 3.674E-13   & 0.000E+00 \\
(14-1) O(6)  &   29663.65 &   29664.29 &   0.3371  & 0.3371 &  8.827E-12 &  8.34E-12& 4.629E-12& 0.000E+00 \\
(14-1) Q(4)&     30846.66  &   30847.35 &   0.3242 &  0.3242 & 1.453E-15  & 2.44E-15 & 3.318E-15 & 6.459E-16 \\
(14-1) S(2)   &  31620.20  &   31620.92 &  0.3163 &    0.3162 &  7.831E-12 &  7.46E-12 & 6.491E-12 &  0.000E+00 \\
(14-2) O(6)  &   25856.91   &  25856.81 &  0.3867   & 0.3867 &  5.045E-11 &   4.76E-11   & 2.627E-11&  0.000E+00  \\
(14-2) Q(4)   &  26978.14   &26978.08   & 0.3707  & 0.3707  & 7.424E-13  & 7.42E-13 & 7.096E-13 & 7.808E-15 \\
(14-2) S(2)    & 27711.68  &  27711.65 &    0.3609  &    0.3609   &  6.921E-11  &  6.62E-11 & 5.802E-11 & 0.000E+00  \\
(14-3) O(6)   &  22278.93  &  22278.15&  0.4489  &  0.4489& 1.788E-10 &  1.71E-11 & 9.437E-11& 0.000E+00  \\
 (14-3) Q(4) &    23339.23    &  23338.49 & 0.4285 &   0.4285 &  1.960E-11  &  1.82E-11 & 1.755E-11 & 5.037E-14  \\
 (14-3) S(2) &     24033.34  &    24032.62 &   0.4161  & 0.4161 &  4.159E-10&  3.94E-10 & 3.458E-10 &  0.000E+00  \\
(14-4) O(6) &    18927.65      & 18926.27 &  0.5283 &   0.5284 &  4.356E-10  & 4.13E-10 & 2.276E-10 & 0.000E+00  \\
 (14-4)  Q(4) &     19927.34&    19925.99 &  0.5018    &  0.5019 &  1.794E-10  & 1.72E-10   &1.660E-10  & 2.697E-13  \\
 (14-4) S(2)  &   20582.30  &  20580.96 &  0.4859   & 0.4859 & 1.843E-09&  1.76E-9 &  1.543E-09 &  0.000E+00  \\
 (14-5) O(6)   &  15803.22 &  15801.29&   0.6328 &   0.6329 &  6.710E-10 &  6.44E-10  & 3.547E-10 & 0.000E+00  \\
 (14-5) Q(4)  &   16742.01  &  16740.11  &   0.5973   &  0.5974 &  1.059E-09&  1.00E-9& 9.695E-10 & 1.122E-12  \\
 (14-5) S(2) &    17357.70   &17355.81   &0.5761  &  0.5762  &  6.535E-09   &  6.21E-9 & 5.449E-09 & 0.000E+00  \\
(14-6) O(6)   &  12908.28 &  12905.89&    0.7747  & 0.7748 & 4.900E-10 &  4.62E-10& 2.547E-10 &   0.000E+00  \\
 (14-6) Q(4)  &   13785.19  &   13782.81& 0.7254  &   0.7255 &  4.521E-09  &  4.31E-9 &  4.162E-09 & 3.697E-12  \\
 (14-6) S(2)   &  14361.03 &14358.67   &  0.6963    & 0.6964 &  1.895E-08    &  1.80E-8 &  1.583E-08 &  0.000E+00  \\
(14-7) O(6) &    10248.53  & 10245.76   & 0.9757  &  0.9760&     6.814E-11&  6.13E-11 & 3.416E-11 & 0.000E+00  \\
 (14-7) Q(4) &    11061.63& 11058.86 &      0.9040 &   0.9042  & 1.611E-08 &  1.53E-8 & 1.477E-08 & 8.560E-12  \\
 (14-7) S(2)   &  11596.54 &  11593.76 &    0.8623  &  0.8625  &    4.700E-08  & 4.46E-10 & 3.919E-08 & 0.000E+00  \\
 (14-8) O(6)    &  7833.35 &   7830.29&    1.2766   &  1.2771  & 8.648E-09  &  8.28E-9   & 4.589E-09 &0.000E+00  \\
 (14-8) Q(4) &     8579.59  &   8576.52&   1.1656 &    1.1660 & 4.628E-08 & 4.41E-8 & 4.265E-08 & 1.649E-11  \\
 (14-8) S(2)    &  9071.74  & 9068.66 &     1.1023 & 1.1027  &   9.460E-08 &  9.01E-8 &  7.916E-08 & 0.000E+00  \\
 (14-9) O(6)   &   5676.83 &   5673.63 &  1.7615  &  1.7625 &  4.647E-08 &  4.40E-8 & 2.436E-08 & 0.000E+00  \\
(14-9) Q(4)  &    6351.62   &   6348.38 &   1.5744 &  1.5752 &    8.297E-08 &  7.86E-8 &7.607E-08 & 8.284E-11  \\
 (14-9) S(2)  &  6798.30    &   6795.03 & 1.4710  &  1.4717 & 1.210E-07  &1.15E-7&  1.008E-07 & 0.000E+00   \\
(14-10)  O(6)  &    3799.29 &  3796.14 &    2.6321&  2.6343 &  5.795E-08&   5.49E-8 & 3.042E-08 & 0.000E+00  \\
(14-10) Q(4)   &   4395.87  &  4392.63 &  2.2749 &  2.2765  &  5.714E-08  & 5.42E-8 & 5.244E-08 & 6.860E-10  \\
(14-10)  S(2) &      4793.09 &   4789.80 &  2.0863  &  2.0878 &  5.673E-08&   5.38E-8  & 4.731E-08 & 0.000E+00  \\
(14-11) O(6)  &    2229.54   & 2226.70 & 4.4852 &  4.4910 & 1.336E-08   &1.26E-8 & 6.997E-09 & 0.000E+00   \\
 (14-11) Q(4)  &    2737.94  & 2734.94  &     3.6524 &  3.6564 &  6.916E-09 &  6.54E-9 & 6.326E-09 & 2.260E-09  \\
 (14-11) S(2) &      3079.90 &   3076.81 &     3.2469  & 3.2501  & 1.796E-09  &  1.70E-9 & 1.484E-09 &0.000E+00  \\
 (14-12) O(6)   &   1008.78 &   1006.64 &  9.9130    &   9.9340 & 1.804E-10 &  1.69E-10  & 9.403E-11 & 0.000E+00  \\
(14-12) Q(4)&      1413.76  & 1411.35  &    7.0734 &   7.0854  &  2.739E-11 &  2.61E-11 &  2.568E-11 & 1.744E-09  \\
 (14-12) S(2)   &   1691.82    &  1689.27 &  5.9108   &  5.9197 & 1.349E-09&  1.28E-9 & 1.130E-09 & 0.000E+00 \\ 
 (14-13) O(6) &      199.07 &    198.21 & 50.2325 &   50.4513 &  1.164E-14 & 1.07E-14 &  6.157E-15 & 0.000E+00  \\
 (14-13) Q(4) &      474.57   &  473.24  &   21.0717 &  21.13074 &  1.053E-11 &  9.82E-12 & 9.661e-12 &1.422E-10  \\
 (14-13) S(2) &      675.05 & 673.47 &    14.8136   &  14.8485  &  1.138E-10 & 1.07E-10 &  9.483E-11 & 0.000E+00  \\
 (14-14) S(2)  &      94.92  &  94.90 &  105.3536 & 105.3695 &  7.280E-15 & 6.71E-15 & 6.049E-15& 0.000E+00 \\ 
\hline
  \end{tabular}}
\tabnote{
 $\tilde{\nu}^{adiab}$, $\tilde{\nu}$ are the transition wavenumbers computed respectively from E  ($v=14, J=4$) = -0.0273  cm$^{-1}$, E ($v=14, J=4$) = 1.2467 cm$^{-1}$ whereas $\lambda^{adiab}$ , $\lambda$   are the corresponding wavelengths. $W^{EQ}$ is the electric quadrupole contribution to the Einstein coefficient from the present computations.  $W^{EQ}_{Forrey}$  and     $W^{EQ}_{\rm{J{\acute{o}}{\acute{z}}wiak}}$ are the values reported from \cite{Forrey:16}
 and calculated from \cite{wcislo:20}. $W^{MD}$
is the magnetic dipole contribution.  The total radiative decay rate of v=4, J=4 is 6.991E-07  s$^{-1}$.}
\label{tab:v14}
\end{table}

However, we note that $J=4$ corresponds to para-H$_2$, that implies that the total nuclear spin is zero, cancelling possible hyperfine couplings and degeneracy.
The hyperfine components of all rovibrational quadrupole transitions  of o-H$_2$  together with the single para-H$_2$ transitions have in fact been computed by \cite{wcislo:20} who use the potential function  of \cite{Komasa:18} and the H2Spectre code of Czachorowski\footnote{Fortran source code, 2019, ver7.0, University of Warsaw, Poland; 2019\\ http://www.qcg.home.amu.edu.pl/qcg/public.html/H2Spectre.html.}.  We have verified that the  $\tilde{\nu}^{adiab}$  values reported in Table \ref{tab:v14} are perfectly corresponding to the transition frequencies reported in \cite{wcislo:20} for the quadrupole emission transitions  arising from $v=14, J=4$  (as well as for the other transitions). However, the electric quadrupole transition emission rates  computed from \cite{wcislo:20} 
are slightly different  from our values and those of \cite{Forrey:16} as displayed in Table \ref{tab:v14}. 
We use the formula 
\begin{equation*}
W_{v'J' \rightarrow v"J"}^{EQ} =  2.797 \times 10^{-23} \cdot \tilde{\nu}^5  \cdot {\rm{Quad. \, Moment}}^2 
\end{equation*}

where $\tilde{\nu} = {\rm{Frequency}}/c$, using the values 'Frequency' and 'Quad. Moment' reported in the supplementary information of \cite{wcislo:20}\footnote{$c$ is the light velocity = 2.99792458e10 cm/s.}.
The values are within less than 20\% for Q and S transitions but they reach a systematic factor of about 2 for the O(6) transitions.  We find similar discrepancies for the O  emission rates of other para-H$_2$ transitions whereas our computations agree very well with those of  Wolniewicz et al. \cite{wolniewicz:98}.

The magnetic dipole contribution to the emission rate is found negligible here compared to the electric quadrupole contribution.
Detection of one of the transitions displayed in Table \ref{tab:v14}
would require a dissociation rate significantly smaller than the total radiative decay rate of 6.99 $\times$ 10$^{-7}$ s$^{-1}$, corresponding to a lifetime larger than  16 days.

The comparison with the computations reported in \cite{Forrey:16} is again found very good for the electric quadrupole emission rates. The only significant discrepancy is obtained in the case of the (14-1) Q(4) electric quadrupole transition emission rate for which a factor close to two is present. However, the extremely low  emission rate value of the order of 10$^{-15}$ s$^{-1}$ and a $\Delta v= 13$ gap explain  the specific sensitivity of the corresponding matrix element to the molecular parameters used in the computations. 

\section{Conclusions}
We propose a new and simple way to compute radiative emission rates involving quasibound state levels where the corresponding eigenfunction is computed as a bound state  trapped in a modified long range potential. We confirm the approach of  Forrey \cite{Forrey:16} and contradict  the assertions of \cite{Perez:19}
in their applications to electric quadrupole transitions in ground state molecular Hydrogen. The procedure is valid for any diatomic system with a $^1\Sigma^+$  X ground state and easily extendable to an open shell system where additional hamiltonian terms should be included. 
We provide accurate wavelengths, wave-numbers and emission rates for both electric quadrupole and magnetic dipole transitions  in H$_2$ thanks
to the highly precise molecular potential function including adiabatic and relativistic corrections \citep{Komasa:18}. This is beautifully confirmed
by the comparison with astrophysical spectra \citep{Pike:16} for the (2-1) S(27) transition at 2.1785 $\mu$m  found towards the shocked region Herbig-Haro 7. The corresponding upper level is quasibound and located 470 cm$^{-1}$ above the dissociation limit of H$_2$ and the corresponding Einstein coefficient is 2.94 $\times$ 10$^{-6}$ s$^{-1}$. 

We  also suggest that the unidentified transition reported at 2.1044 $\mu$m in these same observations could correspond to  (1-0) S(29), predicted at 2.1042 $\mu$m in our computations with a radiative transition emission 
rate of 2.1 $\times$ 10$^{-6}$ s$^{-1}$. The corresponding upper  quasibound level is located  at 1056 cm$^{-1}$ above the dissociation limit in that case.
This is the first time, to our knowledge, that transitions emitted from quasibound levels are  identified in astrophysical environments.
The role of quasibound levels  is well recognized in chemical reactivity \citep{Bain:72,Abgrall:76} but 
there are no study of their  possible collisional excitation  in the literature.
The total radiative decay of that state is 5.22 $\times$ 10$^{-6}$ s$^{-1}$, whereas its dissociation rate is 2.45 $\times$ 10$^{-7}$ s$^{-1}$.
Such an observation could then be used to constrain the lifetime of that excited molecular hydrogen cloud, i.e. about or less than 1/(2.45  $\times$ 10$^{-7}$) s$^{-1}$, 
corresponding to less than 2 months!  The present identifications open a new challenge for studying astrophysical shocked regions.

We finally predict the possible infrared spectrum emitted from $v=14, J=4$, a level for which the quasibound nature is under debate. The para character of that level
implies a total nuclear spin $I=0$, preventing any hyperfine splitting at long range, contrarily to the suggestion of \cite{Selg:11}.

\section*{Acknowledgements}
We thank the referees for useful comments and for driving our attention on the $v=14, J = 4$ level and to the reference \cite{wcislo:20}. We acknowledge support from the Programme National  de Physique et Chimie du Milieu Interstellaire  (PCMI) of CNRS/INSU with INC/INP co-funded by CEA and CNES.

 {}
\bibliographystyle{tfo}

\end{document}